\documentclass[aps,pra,floatfix,showpacs,preprint,onecolumn]{revtex4-2}
\usepackage[utf8]{inputenc}

\usepackage[english]{babel}
\usepackage{latexsym}
\usepackage{indentfirst}
\usepackage{amsmath}
\usepackage{epsfig}
\usepackage{graphicx}
\usepackage{bm}
\usepackage{comment,cancel}
\usepackage{amsfonts,amssymb,latexsym,amsmath,enumerate,amsthm}
\usepackage{indentfirst}
\usepackage{ulem}

\newcommand{\be}{\begin{equation}}
 \newcommand{\ee}{\end{equation}}
\newcommand{\bear}{\be\begin{array}}
\newcommand{\bea}{\begin{eqnarray}}
\newcommand{\eea}{\end{eqnarray}}

\newcommand{\bk}{{\bm k}}

\newcommand{\la}{\langle}
\newcommand{\ra}{\rangle}

\usepackage[dvipsnames]{xcolor}

\newcommand{\dst}{\displaystyle}
\newcommand{\fr}[2]{\frac{{\dst #1}}{{\dst #2}}}

\begin{document}
\title{Emission of twisted photons by a scalar charged particle \\in a strong magnetic field}
\author{D. Karlovets}
\author{A. Di Piazza}
\affiliation{Max Planck Institute for Nuclear Physics, Saupfercheckweg 1, D-69117 Heidelberg, Germany}

\begin{abstract}
We consider the emission of a photon by a scalar charged particle in a constant and uniform magnetic field. In contrast to the conventional approach with both photon and outgoing charge being assumed to be detected, we study the case where only the charge is detected and investigate the properties of the emitted photon. The background magnetic field is taken into account exactly in the calculations and the charge is described by relativistic Landau states. It is shown that the emitted photon state represents a twisted Bessel beam with a total angular momentum given by $\ell-\ell'$, where $\ell$ and $\ell'$ are angular momentum quantum numbers of the initial and final charged particle, respectively. The majority of photons emitted by unpolarized charges, especially in the hard X-ray and $\gamma$-ray range and in critical and sub-critical magnetic fields, as compared to the Schwinger value of $H_c = 4.4\times 10^9$ T, turn out to be twisted with $\ell-\ell'\gtrsim 1$.

\end{abstract}

\maketitle

\section{Introduction}

Classical electrodynamics predicts that an electron moving along a helical path in a magnetic field, in a helical undulator, or in a circularly-polarized laser wave emits radiation in the form of cylindrical waves \cite{Katoh, Katoh2, Epp, Epp2}. Such an electromagnetic field consists of the so-called twisted photons with orbital angular momentum (OAM) \cite{Allen, Torres, And, SerboUFN}. The quantum picture of this process is much more complex due to entanglement between the photon and the electron, and -- depending on a post-selection protocol -- twisted photons may not be emitted even if the classical electron path is helical \cite{EPJC, EPJC2}.
 
The Landau levels of electrons and other charged particles in the magnetic field represent a cornerstone for quantum theory of synchrotron radiation \cite{ST}, which plays a key role in describing such astrophysical objects as the neutron stars \cite{Bagrov}, in operation of modern storage rings, and in producing spin-polarized electron beams. Moreover, the properties of radiation generated by ultrarelativistic charged particles in more complex constant electromagnetic fields can be very similar to those of synchrotron radiation in the magnetic field \cite{ST, Bagrov}. The Landau states \cite{ST, Bagrov, BLP} are also crucially important for many phenomena in solids \cite{Bloch1, Melting, Bloch2} and, in particular, in graphene \cite{graphene1, graphene2, graphene3}, in plasma \cite{Plasma}, including quantum Hall effect \cite{graphene4}, diamagnetism of metals, etc. In addition, these states are used for describing the quantum dynamics of such charged particles as electrons, protons, ions, etc. inside Penning traps and their modifications \cite{Penning0, Penning1, Penning2}. See Ref. \cite{Jentschura_2023} for a recent derivation of the Landau levels by means of an algebraic approach.

Here, we study single photon emission by a spin-zero charge moving along the field lines of a constant and homogeneous magnetic field ${\bm H} = \{0,0,H\}$ of an arbitrary strength $H>0$ \cite{ST, Bagrov}. Since we will assign to the charged particle the numerical values of the charge and of the mass of the electron, for simplicity, we will denote the particle as ``electron''. For an electron with initial four-momentum $p^{\mu}$, the process is controlled by the two Lorentz- and gauge-invariant parameters $b=\sqrt{(F^{\mu\nu})^2}/H_c = \sqrt{2}H/H_c$ and $\chi = \sqrt{-(F^{\mu\nu} p_{\nu})^2}/H_c$, where $F^{\mu\nu}$ is the electromagnetic field tensor and $H_c = 4.4\times 10^9$ T is the Schwinger's (critical) magnetic field strength (the metric tensor is $\eta^{\mu\nu}=\text{diag}(+1 -1 -1 -1)$). In our case the magnetic field is taken into account exactly in the calculations and the electron will be described by relativistic Landau levels with different quantum numbers.

One of our aims is to ascertain whether the emitted photon is twisted, i.e. its total angular momentum (TAM) projection onto the magnetic field axis has a well-defined value. For this reason, we assume that only the final electron is measured and we investigate the resulting photon state \cite{EPJC2}. The difference between this ``evolved'' state and the conventionally used ``detected'' photon state (for instance, in Ref. \cite{BKL, BKL2, Kruining, Mar}) is somewhat analogous to that between the polarization state evolving from the process itself and the detected polarization state that was post-selected by a detector (see, for instance, Ref. \cite{BLP}). The photon evolved state is defined by an $S$-matrix element $S_{fi}$ and, more importantly, the phase of this photon state depends on that of the matrix element, whereas the emission probability only depends on $|S_{fi}|^2$. Thus, this approach also elucidates the role of the $S$-matrix element's phase. Correspondingly, due to the entanglement between the outgoing electron and the emitted photon, the evolved photon state depends on the quantum numbers of the (measured) outgoing electron state.

The system of units $\hbar = c = 1$ is used unless stated otherwise, whereas the electron mass and charge are denoted as $m$ and $e<0$, respectively. The four-vector potentials of the background magnetic field and of the radiation field are taken in the Coulomb gauge.

\section{Evolved photon state: definition}

We want to study the emission $e \to e' + \gamma$ of a photon by an electron in the mentioned magnetic field within the first order of the perturbation theory (we recall that with electron we mean here a scalar particle with mass and charge given by those of the electrons). Thus, by denoting as $\hat{S}^{(1)}$ the first-order $S$-matrix in the Furry picture, with the interaction of the electron with the magnetic field being taken into account exactly, if the system is in the initial state $|i\ra=|i_e\ra$, representing a single electron, the final state is given by $|f\ra=\hat{S}^{(1)}|i_e\ra$. By assuming that the measured final electron state is $|f_e\ra$, then the system collapses into the state
\bea
|f'\ra = P(f_e)\hat{S}^{(1)}|i_e\ra=\sum\limits_{f_{\gamma}} |f_e,f_{\gamma}\ra\la f_e,f_{\gamma}|\hat{S}^{(1)}|i_e\ra,
\eea
where the projector operator
\begin{equation}
P(f_e)=|f_e\ra\la f_e|\otimes\hat{1}_{\gamma} = |f_e\ra\la f_e|\otimes\sum\limits_{f_{\gamma}} |f_{\gamma}\ra \la f_{\gamma}|=\sum\limits_{f_{\gamma}} |f_e,f_{\gamma}\ra \la f_e,f_{\gamma}|
\end{equation}
has been used, with $|f_{\gamma}\ra$ being a complete set of one-particle photon states. By denoting as $S^{(1)}_{fi} = \la f_e,f_{\gamma}|\hat{S}^{(1)}|i_e\ra$ the first-order single-photon emission matrix element and by observing that the electron state $|f_e\ra$ can be taken out from the sum over $f_{\gamma}$, we conclude that after the final electron is detected in the state $|f_e\ra$, the system is in the state $|f'\ra = |f_e\ra\otimes|\gamma\ra_{\text{ev}}$, with
\bea
|\gamma\ra_{\text{ev}} = \sum\limits_{f_{\gamma}} |f_{\gamma}\ra S^{(1)}_{fi},
\eea
being the ``evolved'' photon state \cite{EPJC2}.

Below, for the sake of definiteness, we use the plane waves with the momentum ${\bm k}$ and the helicity $\lambda = \pm 1$ as complete set of one-particle photon states (photons with scalar and longitudinal polarization cannot be produced and ultimately propagate in vacuum) and then we have effectively
\bea
\hat{1}_{\gamma} = \sum\limits_{\lambda = \pm 1}\int\frac{d^3 k}{(2\pi)^3}\frac{1}{2\omega} |{\bm k},\lambda\ra \la {\bm k},\lambda|,
\eea
with $\omega=|\bm{k}|$, such that we need to investigate the $S$-matrix amplitude $S^{(1)}_{fi} = \la f_e;{\bm k}, \lambda|\hat{S}^{(1)}|i_e\ra$ of emission of a plane-wave photon, $e \to e' + \gamma_{{\bm k},\lambda}$. The state $|{\bm k},\lambda\ra$ is defined as $|{\bm k},\lambda\ra=a^{\dag}_{\lambda}(\bm{k})|0\ra$ and it is normalized such that $\la {\bm k}',\lambda'|{\bm k},\lambda\ra=2\omega (2\pi)^3\delta^3(\bm{k}-\bm{k}')\delta_{\lambda\lambda'}$ (the vacuum state $|0\ra$ is normalized to unity). In this way, the photon evolved state is
\bea
&\displaystyle |\gamma\ra_{\text{ev}} = \sum\limits_{\lambda = \pm 1}\int\frac{d^3k}{(2\pi)^3}\frac{1}{2\omega}\, S_{fi}^{(1)}|{\bm k},\lambda\ra.
\eea
The electromagnetic field operator at the spacetime point $(t,\bm r)$ is defined as
\bea
\label{A}
&\displaystyle \hat{\bm{A}}(t,\bm{r})=\sum_{\lambda}\int \frac{d^3\bm{k}}{(2\pi)^3}\frac{1}{\sqrt{2\omega}}\left[\bm{A}_{\lambda}(t,\bm{r};{\bm k})\hat{a}_{\lambda}(\bm{k})+\bm{A}^*_{\lambda}(t,\bm{r};{\bm k})\hat{a}^{\dag}_{\lambda}(\bm{k})\right],\cr
& \displaystyle \bm{A}_{\lambda}(t,\bm{r};{\bm k}) = \frac{\bm{e}_{\lambda}(\bm{k})}{\sqrt{2\omega}} e^{-i(\omega t-\bm{k}\cdot\bm{r})},
\eea
with the polarization vector $\bm{e}_{\lambda}(\bm{k})$ being specified below.

\section{Relativistic Landau state}

In the constant and homogeneous magnetic field ${\bm H} = \{0,0,H\}, H > 0,$ with the four-vector potential
\bea
\mathcal A^{\mu} = \frac{H}{2}\{0,-y, x, 0\},
\label{AH}
\eea
a charged spinless particle with mass $m$ and charge $e<0$ is described by the Klein-Gordon equation 
\bea
\left((i\partial^{\mu} - e \mathcal A^{\mu})^2 - m^2\right) \Psi(x) = 0.
\label{KG}
\eea
The characteristic scales of the problem are
\begin{align}
H_c &= \frac{m^2}{|e|}= 4.4\times 10^9\, \text{T} - \text{the critical magnetic field},\\
\lambda_c &= \frac{1}{m}= 3.8 \times 10^{-11}\,\text{cm} - \text{the electron Compton wavelength},\\
\rho_H &= \sqrt{\frac{4}{|e| H}} = 2\lambda_c \sqrt{\frac{H_c}{H}} - \text{the rms-radius of the ground state (see below) times $\sqrt{2}$}.
\end{align}
In cylindrical coordinates, Eq. (\ref{KG}) has the following exact stationary solution (called relativistic Landau state) \cite{ST}:
 \bea
& \displaystyle \Psi_i(t,\bm{x})= N_i \left(\frac{\rho}{\rho_H}\right)^{\ell} L_{s}^{\ell}\left(2\rho^2/\rho_H^2\right)\, \exp\left\{-it\varepsilon + ip_z z + i\ell\phi_r - \rho^2/\rho_H^2\right\},\cr & \displaystyle \varepsilon = \sqrt{m^2 + p_z^2 + p_{\perp}^2},\ p_{\perp}^2 = \frac{8}{\rho_H^2} (s + \ell +1/2) = 2m^2 \frac{H}{H_c} (s + \ell +1/2),
  \label{RDau}
 \eea
where $L_{s}^{\ell}$ are the associated (generalized) Laguerre polynomials \cite{GRR}. The independent quantum numbers are the continuous longitudinal momentum $p_z$ and the two discrete quantum numbers $\ell$ and $s$, which describe the transverse motion of the charge. In addition, it is
\be
s \geq 0,\ s + \ell \geq 0,\ \text{such that}\ \ell \geq -s,
\ee
where $s +1 = 1,2,3,...$ defines the number of radial maxima of the probability density, and $\ell = 0, \pm 1, \pm 2, ...$ is the canonical angular momentum (cf. with the non-relativistic case in Ref. \cite{NJP})
\bea
& \displaystyle \la\hat{L}_z^{\text{can}}\ra = \frac{\varepsilon}{m} \int d^3x\, \Psi_i^*(t,\bm{x})\hat{L}_z^{\text{can}}\Psi_i(t,\bm{x}) = \ell,\cr
& \displaystyle \hat{L}_z^{\text{can}} = [\hat{\bm r} \times \hat{\bm p}^{\text{can}}]_z = -i\frac{\partial}{\partial \phi_r}.
\eea

Note that the choice of the potential (\ref{AH}) is not unique and taking $\mathcal A^{\mu} = \{0,-yH, 0, 0\}$ one also obtains ${\bm H} = \{0,0,H\}$. The eigenfunctions of the latter problem, however, are Hermite-Gaussian states with no OAM \cite{LL3, ST} rather than the above Laguerre-Gaussian ones. Thus, the analysis of the radiation for an electron initially in a Laguerre-Gaussian state is simplified in the gauge (\ref{AH}). Our assumption, in fact, is that in the far past (at $t \to -\infty$) the electron was produced in the field (say, emitted from an atom) in a cylindrically symmetric Laguerre-Gaussian state. If the electron is produced outside of the field, still a smooth transition of a freely-propagating Laguerre-Gaussian state to the above Landau one (\ref{RDau}) is possible but a detailed description of the transmission process requires a separate analysis \cite{PRA22, Arxiv23}, which is beyond the scope of this paper.

The above Laguerre-Gaussian states are normalized via the following condition \cite{ST}
 \bea
& \displaystyle \frac{\varepsilon}{m} \int d^3x |\Psi_i(t,\bm{x})|^2 = 1
  \label{Norm}
 \eea
and then we find (Eq. (7.414) in Ref. \cite{GRR})
 \bea
& \displaystyle N_i = \sqrt{\frac{m}{\varepsilon}}\, \frac{1}{\sqrt{L\, \pi \rho_H^2}}\,2^{(\ell+1)/2} \sqrt{\frac{s!}{(s+\ell)!}}.
  \label{Normconst}
 \eea
An effective volume of a normalization cylinder is $L\, \pi \rho_H^2$ where $L$ is its height and $\rho_H$ characterizes the mean radius as (see Eq. (19.29) in Ref. \cite{ST})
\be
\la \rho^2\ra = \frac{\varepsilon}{m} \int d^3x\, \Psi_i^*(t,\bm{x}) \rho^2 \Psi_i(t,\bm{x}) = \frac{\rho_H^2}{2} \left(2s + \ell + 1\right).
\label{re2}
\ee
Therefore, the rms-radius of the ground (Gaussian) state with $s= \ell = 0$ is $\sqrt{\la \rho^2\ra} = \rho_H/\sqrt{2}$. For magnetic field strengths of $H \sim 0.1 - 10$ T, as available in terrestrial laboratories, we find $\rho_H \sim 10-100$ nm. However, for the intense fields $H \sim H_c$ that can take place in the neutron stars the corresponding radii are of the order of the electron Compton wave length itself, $\rho_H \sim \lambda_c$, and the root-mean-square radius of the Landau state scales as $\sqrt{\la\rho^2\ra} \propto \sqrt{2s + \ell + 1}$. As a result, the condition of the spatially homogeneous field seems to work even better for intense fields, even if the quantum numbers $s,\ell$ are very large.

The transverse motion of the electron is non-relativistic, $p_{\perp} \ll m$, when 
\bea
\frac{H}{H_c} \left(s + \ell + 1/2\right) \ll 1,\ \text{or}\ s + \ell + 1/2 \ll \frac{H_c}{H}.
\eea

Importantly, the centroid of the Landau states moves only along the $z$ axis, because
\bea
& \displaystyle \la \hat{x}\ra = \la \hat{y}\ra = \la \hat{p}_x\ra = \la \hat{p}_y\ra = 0,\cr
& \displaystyle \la \hat{p}_z\ra = p_z.
\eea
Thus, although the probability current has an azimuthal component \cite{BliokhPRX}, there is no classical rotation of the wave packet's centroid, because the Lorentz force along the trajectory of the centroid vanishes. Below, we assume that $p_z\ge 0$.

The Landau states resemble the freely propagating vortex electrons with the quantized angular momentum but without the magnetic field \cite{Bliokh0, Bliokh, BliokhPRX, McMorran, Ivanov, NJP}. In particular, the Laguerre-Gaussian beams of the twisted electron states look very similar to Eq. (\ref{RDau}) \cite{Bliokh, BliokhPRX} and can also exactly obey the quantum wave equations \cite{NJP}.

\section{Matrix element for photon emission}

In scalar QED, the first-order transition matrix element with emission of a plane-wave photon described by the state $|{\bm k},\lambda\ra$ is
 \bea
& \displaystyle S^{(1)}_{fi} = -ie\int d^4x\, j_{fi}^{\mu}(t,\bm{x})e_{\lambda,\mu}^*(\bm{k})\,e^{i(\omega t-\bm{k}\cdot\bm{r})},\cr
& \displaystyle j_{fi}^{\mu}(t,\bm{x}) = \frac{1}{2m}\Psi_f^*(t,\bm{x})(i\partial^{\mu} -e\mathcal A^{\mu})\Psi_i(t,\bm{x})+ \frac{1}{2m}\Psi_i(t,\bm{x})(i\partial^{\mu} -e\mathcal A^{\mu})^*\Psi_f^*(t,\bm{x}),
  \label{Sfi}
 \eea
 where $\mathcal A^{\mu}$ is the magnetic field four-potential in Eq. (\ref{AH}), where $e_{\lambda}{}^{\mu}(\bm{k})=(0,\bm{e}_{\lambda}(\bm{k}))$, with $\bm{k}\cdot\bm{e}_{\lambda}(\bm{k})=0$ is the polarization four-vector of the photon (the index $\lambda$ is clearly not a Lorentz index and, since below we will use a three-dimensional notation, there will be then no possibility of confusion), and where
 \bea
 \bm{k} = \omega\{\sin\theta\cos\phi, \sin\theta\sin\phi, \cos\theta\}
 \eea
 is the photon wave vector. We will also need the photon's transverse momentum
 \bea
k_{\perp} = \omega \sin\theta.
\eea 

The final (detected) Landau state of the emitting electron is
 \bea
& \displaystyle \Psi_f(t,\bm{x}) = N_f \left(\frac{\rho}{\rho_H}\right)^{\ell'} L_{s'}^{\ell'}\left(2\rho^2/\rho_H^2\right)\, \exp\left\{-it\varepsilon' + ip'_z z + i\ell'\phi_r - \rho^2/\rho_H^2\right\},\cr & \displaystyle \varepsilon' = \sqrt{m^2 + (p_z')^2 + (p'_{\perp})^2},\ (p'_{\perp})^2 = \frac{8}{\rho_H^2} (s' + \ell' +1/2) = 2m^2 \frac{H}{H_c} (s' + \ell' +1/2),\cr
& \displaystyle N_f = \sqrt{\frac{m}{\varepsilon'}}\, \frac{1}{\sqrt{L\, \pi \rho_H^2}}\,2^{(\ell'+1)/2} \sqrt{\frac{s'!}{(s'+\ell')!}}.
  \label{RDauF}
 \eea
It has a radial index $s'$, a canonical angular momentum $\ell'$, which not necessarily coincide with $s$ and $\ell$, respectively.
The emitted photon polarization vector $\bm{e}_{\lambda}(k)$ can be taken, for instance, in a basis of the circularly polarized states with the helicity 
\be
\lambda = \pm 1.
\ee
In the Coulomb gauge with $\bm k \cdot \bm e_{\lambda}(\bm{k}) = 0$, one can choose the photon polarization vectors as \cite{SerboUFN}
 \bea
&\bm e_{\lambda}(\bm{k}) = \sum\limits_{\sigma = 0,\pm 1}\exp\{-i\sigma\phi\}d^{(1)}_{\sigma\lambda}(\theta)\bm \chi_{\sigma},\cr
& \bm \chi_0 = (0,0,1),\ \bm \chi_{\pm 1} = \mp\frac{1}{\sqrt{2}}(1,\pm i,0),
  \label{ee}
 \eea
where $d^{(1)}_{\sigma\lambda}(\theta)$ are the small Wigner functions \cite{Varsh}
\bea
\label{ekl}
& \displaystyle d_{\lambda \lambda'}^{(1)}(\theta) =\fr 12 \left(1+\lambda \lambda'\cos\theta\right),\ d_{11}^{(1)} = \cos^2(\theta/2),\, d_{1-1}^{(1)} = \sin^2(\theta/2),\cr
& d_{\lambda 0}^{(1)}(\theta) =-d_{0 \lambda}^{(1)}(\theta)=-\fr{\lambda}{\sqrt{2}} \sin\theta,
d_{00}^{(1)}(\theta)=\cos\theta,\\
& \sum\limits_{\sigma = 0,\pm 1} d_{\sigma_1\sigma}^{(1)}(\theta)\,d_{\sigma_2\sigma}^{(1)}(\theta) = \delta_{\sigma_1\sigma_2}.
\label{d}
\eea
The vectors $\bm \chi_{\sigma}$ represent eigenvectors for the photon spin operator
\begin{equation}
\hat{s}_z=\begin{pmatrix}
0 & -i & 0\\
i & 0 & 0 \\
0 & 0 & 0 
\end{pmatrix},
\end{equation}
with the eigenvalues $\sigma = 0,\pm 1$,
\be
\hat{s}_z \bm \chi_{\sigma} = \sigma \bm \chi_{\sigma}.
\ee
Note that the helicity state from Eq. (\ref{ee}) has a vanishing $z$-projection of the TAM,
\bea
& \displaystyle \hat{\lambda} {\bm e}_{\lambda}(\bm{k}) = \lambda {\bm e}_{\lambda}(\bm{k}),\ \hat{\lambda} = {\bm k}\cdot \hat{{\bm s}}/\omega,\cr 
& \displaystyle \hat{j}_z {\bm e}_{\lambda}(\bm{k}) = 0,\ \hat{j}_z = \hat{L}_z + \hat{s}_z,
\label{jz0}
\eea
where $\hat{L}_z=-i\partial/\partial \phi$. This can be easily seen at $\theta \to 0$:
\bea
& {\bm e}_{\lambda}(\bm{k}) \to \bm \chi_{\lambda} e^{-i\lambda\phi},
\eea
with a somewhat redundant dependence on $\phi$. However, the evolved photon state does not depend on the general phase of $\bm e_{\lambda}(\bm{k})$ because it depends on the combination $|f_{\gamma}\ra\la f_{\gamma}|$.

Now, we regroup the terms in the transition four-current in Eq. (\ref{Sfi}) as follows (recall that $e<0$):
 \bea
& e_{\lambda,\mu}^*(\bm{k})(\partial^{\mu} + i e \mathcal A^{\mu}) = d^{(1)}_{0\lambda}(\theta)\partial_z + \frac{1}{\sqrt{2}}\, \left(d^{(1)}_{-1\lambda}(\theta)e^{-i(\phi-\phi_r)} - d^{(1)}_{1\lambda}(\theta)e^{i(\phi-\phi_r)}\right)\partial_{\rho} + \cr
& + \frac{1}{\sqrt{2}}\frac{i}{\rho}\left(d^{(1)}_{-1\lambda}(\theta)e^{-i(\phi-\phi_r)} + d^{(1)}_{1\lambda}(\theta)e^{i(\phi-\phi_r)}\right)\partial_{\phi_r} - \frac{|e|H\rho}{2\sqrt{2}}\left(d_{1\lambda}^{(1)}(\theta) e^{i (\phi - \phi_r)} + d_{-1\lambda}^{(1)}(\theta) e^{-i (\phi - \phi_r)}\right)  \cr
& \equiv \sum\limits_{\sigma=0,\pm 1} d_{\sigma\lambda}^{(1)}(\theta)\, e^{i\sigma (\phi - \phi_r)} \hat{X}_{\sigma},
  \label{eed}
 \eea
 with
 \begin{align}
 \hat{X}_{+1}&=\frac{1}{\sqrt{2}}\left(\frac{i}{\rho}\partial_{\phi_r} - \partial_{\rho} - \frac{|e|H\rho}{2}\right),\\
 \hat{X}_{-1}&=\frac{1}{\sqrt{2}}\left(\frac{i}{\rho}\partial_{\phi_r} + \partial_{\rho} - \frac{|e|H\rho}{2}\right),\\
 &\hat{X}_{0}=\partial_z.
 \end{align}

Taking into account that $\partial_{\rho}L_s^{\ell}(\rho) = - L_{s-1}^{\ell+1}(\rho)$ for $s \geq 1$, we arrive at 
 \bea
& \displaystyle e_{\lambda,\mu}^*(\bm{k})j^{\mu}_{fi}(t,\bm{r}) = \frac{1}{2m}\Psi_f^*(t,\bm{r}) \Psi_i(t,\bm{r}) \sum\limits_{\sigma=0,\pm 1} e^{i\sigma(\phi - \phi_r)} d_{\sigma\lambda}^{(1)}(\theta)\, X_{\sigma},  
  \label{eedpsil}
 \eea
with $X_{\sigma}$ being given by
\begin{align}
&X_{+1}=i\frac{\sqrt{2}}{\rho_H} \left[2\tilde{\rho} \left(\frac{L_{s-1}^{\ell + 1}}{L_{s}^{\ell}} - \frac{L_{s'-1}^{\ell'+1}}{L_{s'}^{\ell'}}\right) - \frac{\ell}{\tilde{\rho}} - 2 \tilde{\rho}\right],\\
&X_{-1}=-i\frac{\sqrt{2}}{\rho_H} \left[2\tilde{\rho} \left(\frac{L_{s-1}^{\ell + 1}}{L_{s}^{\ell}} - \frac{L_{s'-1}^{\ell'+1}}{L_{s'}^{\ell'}}\right) + \frac{\ell'}{\tilde{\rho}} + 2 \tilde{\rho} \right],\\
&X_0=-p_z-p'_z,
\end{align}
where the argument 
\be
2\tilde{\rho}^2 \equiv 2\rho^2/\rho_H^2
\ee
of the Laguerre polynomials is omitted and the terms containing $L_{s-1}^{\ell+1}$ and $L_{s'-1}^{\ell'+1}$ are non-vanishing only when $s, s' \geq 1$ because $L_0^{\ell}(\rho) = 1$.

The azimuthal integral in Eq. (\ref{Sfi}) can taken by employing the following definition of the Bessel functions:
 \bea
& \displaystyle \int\limits_0^{2\pi}\,\frac{d\phi_r}{2\pi}\, e^{i \ell \phi_r - ix\cos \phi_r} = i^{-\ell} J_{\ell}(x),\ \int\limits_0^{2\pi}\,\frac{d\phi_r}{2\pi}\, e^{i \ell \phi_r + ix\cos \phi_r} = i^{\ell} J_{\ell}(x).
\label{JB}
 \eea
Thus, we arrive at the following matrix element
 \bea
& \displaystyle S^{(1)}_{fi} = -ie N_i N_f\frac{\rho_H}{2m}\, (2\pi)^3 \delta (\omega + \varepsilon' - \varepsilon) \delta (p_z - p_z' - k_z) e^{i(\ell-\ell')\phi} \sum\limits_{\sigma=0,\pm 1} i^{\sigma-\ell+\ell'}\,d_{\sigma\lambda}^{(1)}(\theta)\, \mathcal I_{\sigma}(y),\cr
& \displaystyle  \mathcal I_{\sigma} (y) =\rho_H\int\limits_0^{\infty} d\tilde{\rho}\,\tilde{\rho}^{\ell+\ell'+1}\, X_{\sigma}(\tilde{\rho})\, L_{s}^{\ell}(2\tilde{\rho}^2)L_{s'}^{\ell'}(2\tilde{\rho}^2)\,J_{\ell-\ell' - \sigma}(y \tilde{\rho})\, e^{-2\tilde{\rho}^2},
  \label{Sej}
\eea
where we have denoted
\be
y \equiv k_{\perp}\rho_H.
\ee
The integrals are evaluated with the help of Eq. (20.17) in Ref. \cite{ST} as follows (see details in the Appendix):
\bea
&& \displaystyle \mathcal I_{0}(y) = -\rho_H (p_z+p_z')\, \mathcal F_{s,s'}^{\ell,\ell'}(y),\cr
&& \displaystyle 
\mathcal I_{+1}(y) = -i\sqrt{2}\left(2 \mathcal F_{s,s'}^{\ell,\ell'+1}(y) + (s+\ell) \mathcal F_{s,s'}^{\ell-1,\ell'}(y)\right),\cr
&& \displaystyle
  \mathcal I_{-1}(y) = -i\sqrt{2}\left(2 \mathcal F_{s,s'}^{\ell+1,\ell'}(y) + (s'+\ell') \mathcal F_{s,s'}^{\ell,\ell'-1}(y)\right),\cr
&& \displaystyle \mathcal F_{s,s'}^{\ell,\ell'}(y) = \int\limits_0^{\infty} dx\,x^{\ell+\ell'+1}\, L_{s}^{\ell}(2x^2)L_{s'}^{\ell'}(2x^2)\,J_{\ell-\ell'}(y x)\, e^{-2x^2} = \cr 
&& \displaystyle = \frac{(s'+\ell')!}{s!}\frac{1}{2^{3(s-s') + 2\ell - \ell' + 2}}\, y^{2(s-s') + \ell-\ell'}\, L_{s'+\ell'}^{s-s'+\ell-\ell'}\left(y^2/8\right)L_{s'}^{s-s'}\left(y^2/8\right)\, e^{-y^2/8}.
\label{Is}
 \eea
Thus, the matrix element contains the damping exponential factor 
\bea
\exp\left\{-\frac{(k_{\perp}\rho_H)^2}{8}\right\} = \exp\left\{-\frac{k_{\perp}^2}{2m^2}\frac{H_c}{H}\right\}.
\label{expk}
\eea

\section{Evolved photon state: analysis}

According to the above analysis the evolved photon state is given by
\be
\begin{split}
\label{gamma_ev}
|\gamma\ra_{\text{ev}} &=  -ie \sum\limits_{\lambda = \pm 1}\int\frac{d^3 k}{(2\pi)^3}\,\frac{N_i N_f}{2\omega }\,\frac{\rho_H}{2m}\, (2\pi)^3 \delta (\omega + \varepsilon' - \varepsilon) \delta (p_z - p_z' - k_z) e^{i(\ell-\ell')\phi} \\
&\quad\times\sum\limits_{\sigma=0,\pm 1} i^{\sigma-\ell+\ell'}\,d_{\sigma\lambda}^{(1)}(\theta)\, \mathcal I_{\sigma}(y)|{\bm k},\lambda\ra.
\end{split}
\ee
By writing the integral in cylindrical coordinates and by noticing that
\be
\delta (\omega + \varepsilon' - \varepsilon) = \delta \left(\sqrt{k_{\perp}^2 + (p_z - p_z')^2} + \varepsilon' - \varepsilon\right) =  \frac{\varepsilon - \varepsilon'}{\kappa} \delta\left(k_{\perp} - \kappa\right),
\label{dF}
\ee
where
\bea
\kappa = \sqrt{(\varepsilon - \varepsilon')^2 - (p_z - p_z')^2} \geq 0, 
\label{kapp}
\eea
the corresponding two integrals in $k_z$ and $k_{\perp}$ can be taken and the result is
\be
\label{gamma_ev_2}
|\gamma\ra_{\text{ev}} =  -ie\rho_H\frac{N_i N_f}{2m\kappa}\sqrt{\frac{\varepsilon-\varepsilon'}{4\omega}} \sum\limits_{\lambda = \pm 1}\int_0^{2\pi}d\phi\, e^{i(\ell-\ell')\phi} \sum\limits_{\sigma=0,\pm 1} i^{\sigma-\ell+\ell'}\,d_{\sigma\lambda}^{(1)}(\theta)\, \mathcal I_{\sigma}(y)|{\bm k},\lambda\ra.
\ee
In this expression, the following equalities are understood:
\begin{align}
y&=\kappa\rho_H,\\
\theta&=\arctan\left(\frac{\kappa}{p_z-p'_z}\right),\\
\bm{k}&=\{\kappa\cos\phi,\kappa\sin\phi,p_z-p'_z\},
\end{align}
such that in the integrand of Eq. (\ref{gamma_ev_2}) only the exponential $\exp[i(\ell-\ell')\phi]$ and the state $|{\bm k},\lambda\ra$ depend on $\phi$.

The expansion of the state $|\gamma\ra_{\text{ev}}$ in Eq. (\ref{gamma_ev}) already shows that it is a so-called Bessel beam. Indeed, it has a definite energy $\omega = \varepsilon - \varepsilon'$, a definite longitudinal momentum $k_z = p_z - p_z'$, and a definite transverse momentum $\kappa$ but not an azimuthal component of the momentum, which is a hallmark of twisted states. This can be more easily recognized by considering the matrix element $\la 0|\hat{\bm{A}}(t,\bm{r})|\gamma\ra_{\text{ev}}$ of the electromagnetic field operator $\hat{\bm{A}}(t,\bm{r})$ [see Eq. (\ref{A})], which can be written as
\begin{equation}
\begin{split}
\la 0|\hat{\bm{A}}(t,\bm{r})|\gamma\ra_{\text{ev}}&=-ie \sum\limits_{\lambda = \pm 1}\int\frac{d^3 k}{(2\pi)^3}\,\frac{N_i N_f}{2\omega }\,\frac{\rho_H}{2m}\, (2\pi)^3 \delta (\omega + \varepsilon' - \varepsilon) \delta (p_z - p_z' - k_z) e^{i(\ell-\ell')\phi} \\
&\quad\times\sum\limits_{\sigma=0,\pm 1} i^{\sigma-\ell+\ell'}\,d_{\sigma\lambda}^{(1)}(\theta)\, \mathcal I_{\sigma}(y)\bm{e}_{\lambda}(\bm{k})e^{-i(\omega t-\bm{k}\cdot\bm{r})}.
\label{ArEv}
\end{split}
\end{equation}
In this way, following Ref. \cite{BLP}, we can interpret the vector
\bea
& \displaystyle {\bm A}^{\text{(ev)}}({\bm k}) = -ie i^{-\ell+\ell'} N_i N_f\,\frac{\rho_H}{2m}\, (2\pi)^3 \delta (\omega + \varepsilon' - \varepsilon) \delta (p_z - p_z' - k_z) e^{i(\ell-\ell')\phi}\cr
& \displaystyle \times \sum\limits_{\sigma=0,\pm 1} i^{\sigma} \mathcal I_{\sigma} \sum\limits_{\lambda=\pm 1} d^{(1)}_{\sigma \lambda}(\theta)\, {\bm e}_{\lambda}(\bm{k}),
\cr & \displaystyle
\bk \cdot {\bm A}^{\text{(ev)}}({\bm k}) = 0,
\label{Akev}
\eea
as the coefficient of the vector potential of the emitted photon in the momentum representation, which is a superposition of the two helicity states. In the limit $\theta \to 0$ we have that $d^{(1)}_{\sigma \lambda}(\theta) \to \delta_{\sigma\lambda}$, so the value $\sigma=0$ does not contribute for very small transverse momenta. By using the representation of the unit polarization vector ${\bm e}_{\lambda}(\bm{k})$ from Eq. (\ref{ee}) and the completeness relation for the small Wigner functions from Eq. (\ref{d}), one can rewrite the vector part as a single sum as follows:
\bea
& \displaystyle \sum\limits_{\sigma=0,\pm 1} i^{\sigma} \mathcal I_{\sigma}\sum\limits_{\lambda=\pm 1} d^{(1)}_{\sigma \lambda}(\theta)\, {\bm e}_{\lambda}(\bm{k}) = \sum\limits_{\sigma=0,\pm 1} i^{\sigma} \mathcal I_{\sigma} \left ({\bm\chi}_{\sigma} e^{-i\sigma\phi} - d^{(1)}_{\sigma 0}(\theta)\,\sum\limits_{\sigma'=0,\pm 1}d^{(1)}_{\sigma' 0}(\theta){\bm\chi}_{\sigma'} e^{-i\sigma'\phi}\right) \cr
& \displaystyle = \sum\limits_{\sigma=0,\pm 1} i^{\sigma} \mathcal I_{\sigma} \left ({\bm\chi}_{\sigma} e^{-i\sigma\phi} - d^{(1)}_{\sigma 0}(\theta)\, {\bm n}\right),
\label{AvecP}
\eea
where we have used that 
\bea
{\bm n} = {\bm k}/\omega = \sum\limits_{\sigma'=0,\pm 1}d^{(1)}_{\sigma' 0}(\theta){\bm\chi}_{\sigma'} e^{-i\sigma'\phi}.
\eea
Thus, the last term in Eq. (\ref{AvecP}) is due to the Coulomb gauge. 

As $\hat{j}_z {\bm e}_{\lambda}(\bm{k}) = 0$, we finally obtain the photon TAM projection
\bea
& \hat{j}_z{\bm A}^{\text{(ev)}}({\bm k}) = (\ell - \ell'){\bm A}^{\text{(ev)}}({\bm k}).
\label{jzAk}
\eea
Importantly, the photon TAM does not depend on the radial quantum numbers $s,s'$ of the electron, which can also change during the emission.

The condition of the positive transverse momentum (\ref{kapp}) defines the interval of the allowed values of the final electron longitudinal momentum $p_z'$. First, we notice that, since $\varepsilon-\varepsilon'=\omega\ge 0$, the condition $\kappa\ge 0$ implies that $\varepsilon-\varepsilon'\ge |p_z-p'_z|$ and then that $p_{\perp}^2 - (p'_{\perp})^2\ge 2[\varepsilon'-\text{sgn}(p_z- p'_z)p'_z]|p_z-p'_z|\ge 0$. Therefore, by solving the equation $\kappa = 0$ with respect to $p_z'$ with given $s'$ and $\ell'$, we find that
\be
p'_z \in \Big [p'_{z,\text{min}}, p'_{z,\text{max}}\Big]= \Big [p_z- \frac{p_{\perp}^2 - (p'_{\perp})^2}{2(\varepsilon-p_z)},p_z+ \frac{p_{\perp}^2 - (p'_{\perp})^2}{2(\varepsilon+p_z)}\Big ], 
\label{pzpint}
\ee
so the width of the interval where the final longitudinal electron momentum lies is 
\be
p'_{z,\text{max}} - p'_{z,\text{min}} = \varepsilon\, \frac{p_{\perp}^2 - (p'_{\perp})^2}{m^2 + p_{\perp}^2} = \varepsilon\, \frac{s + \ell - s' - \ell'}{s+\ell+1/2 + \frac{1}{2}\frac{H_c}{H}} \ge 0,
\label{deltap}
\ee
whereas the average of the possible values of $p'_z$ is
\be
\frac{p'_{z,\text{min}}+p'_{z,\text{max}}}{2} = p_z\left[1-\frac{p_{\perp}^2 - (p'_{\perp})^2}{2(m^2 + p_{\perp}^2)}\right]=p_z\left[1-\frac{s + \ell - s' - \ell'}{2(s+\ell)+1 + \frac{H_c}{H}}\right].
\ee
If $p_{\perp} \ll m, p'_{\perp} \ll m$, then
\bea
& \displaystyle \frac{p'_{z,\text{max}} - p'_{z,\text{min}}}{\varepsilon} \approx 2 \frac{H}{H_c} (s + \ell - s'- \ell'). 
\eea
The interval $p'_{z,\text{max}} - p'_{z,\text{min}}$ grows for larger and larger initial electron energies. 

\begin{figure}[t]
\center{\includegraphics[width=0.99\linewidth]{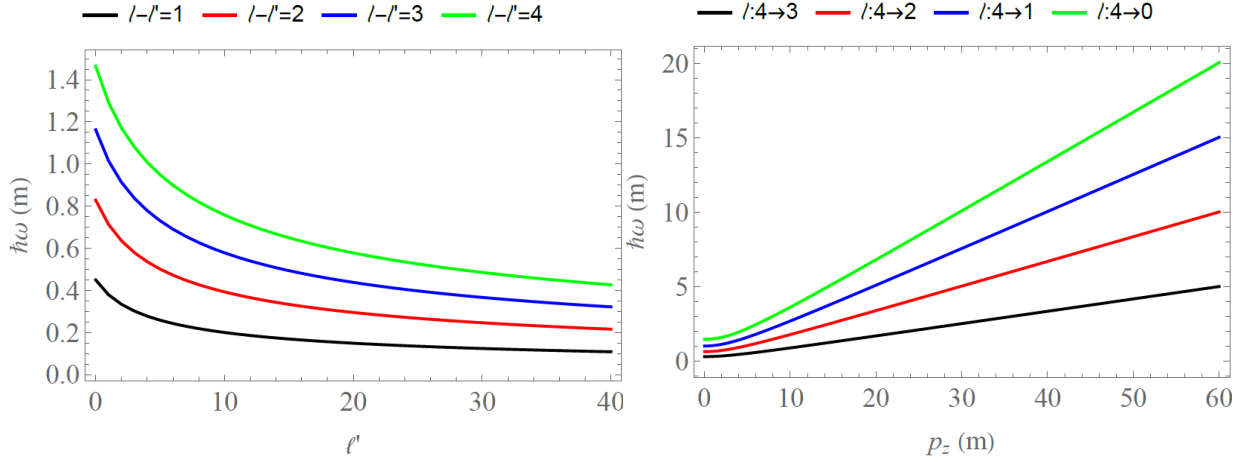}}
\caption{Emitted photon energies in units of the electron rest energy $m = 0.511$ MeV as functions of the final electron quantum number $l'$ (left) and of the initial electron longitudinal momentum $p_z$ (right) for $H=H_c$, $s=s'=1$, $p_z = 10^{-3}\,m$ (left). In general, the transition $l:i\to j$ means $\ell=i$, and $\ell'=j$. The photon TAM equals $\ell - \ell'$ according to Eq. (\ref{jzAk}), and everywhere $p'_z = (p'_{z,\text{min}} + p'_{z,\text{max}})/2$.}
\label{FigSpectrumA}
\end{figure}

\begin{figure}[t]
\center{\includegraphics[width=0.99\linewidth]{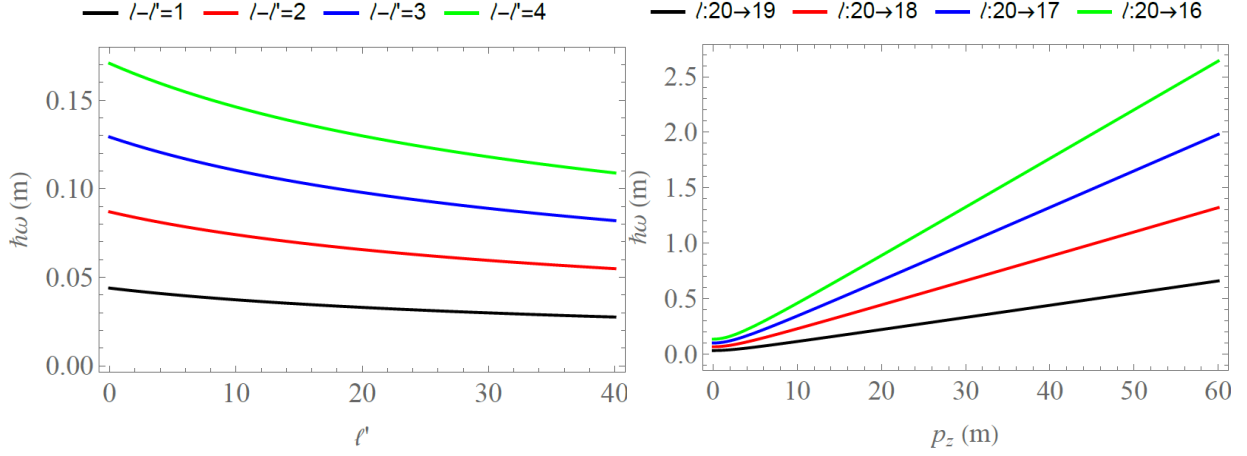}}
\caption{The same as in Fig. \ref{FigSpectrumA}, but for $H=0.1\,H_c$, $s=s'=20$.}
\label{FigSpectrumB}
\end{figure}

As shown in Figs. \ref{FigSpectrumA} and \ref{FigSpectrumB}, the energy of the emitted photon simply coincides with that of synchrotron radiation (see, e.g., Refs. \cite{ST, Bagrov} with the substitutions $\sin \theta = \kappa/\sqrt{\kappa^2 + k_z^2}$ and $\cos \theta = k_z/\sqrt{\kappa^2 + k_z^2}$) with the hard X-ray and $\gamma$-ray photons being emitted in the strong magnetic fields, especially by relativistic electrons.

As we have mentioned, Eq. (\ref{Akev}) is a superposition of two helicity state. For a photon emitted almost along the magnetic field, i.e., for $\theta \to 0, y \to 0$, we have that (recall that $s + \ell \geq 0$ and $s\geq 0$) \cite{GRR}
\bea
& \displaystyle L_{s}^{\ell}(0) = \begin{pmatrix}
s+\ell\\
s
\end{pmatrix}.
\eea
By using this finding, we arrive at
\bea
&& \displaystyle \mathcal F_{s,s'}^{\ell,\ell'}(0) = \delta_{s,s'}\delta_{\ell,\ell'}\frac{1}{2^{\ell+2}} \frac{(s+\ell)!}{s!},\cr
&& \displaystyle \mathcal I_{+1}(0) = \frac{-i\sqrt{2}}{2^{\ell+1}}\delta_{s,s'} \frac{(s+\ell)!}{s!} \left(\delta_{\ell,\ell'+1} + \delta_{\ell-1,\ell'}\right),\cr
&& \displaystyle \mathcal I_{-1}(0) = \frac{-i\sqrt{2}}{2^{\ell+2}}\delta_{s,s'} \frac{(s+\ell + 1)!}{s!} \left(\delta_{\ell+1,\ell'} + \delta_{\ell,\ell'-1}\right).
\eea
As a result, we have in Eq. (\ref{Akev})
\bea
& \displaystyle \sum\limits_{\sigma=0,\pm 1} i^{\sigma} \mathcal I_{\sigma} \left (d^{(1)}_{\sigma 1}(\theta)\, {\bm e}_{+1}(k) + d^{(1)}_{\sigma, -1}(\theta)\, {\bm e}_{-1}(k) \right) = \frac{\sqrt{2}}{2^{\ell + 1}}\frac{\delta_{s,s'}}{s!}\cr
& \displaystyle \times  \Big((s+\ell)!\left(\delta_{\ell,\ell'+1} + \delta_{\ell-1,\ell'}\right){\chi}_{+1} e^{-i\phi} - \frac{1}{2} (s+\ell + 1)!\left(\delta_{\ell+1,\ell'} + \delta_{\ell,\ell'-1}\right){\chi}_{-1} e^{i\phi} \Big).
\label{Apar}
\eea
It is clearly not a helicity state,
\bea
& \displaystyle \hat{s}_z {\bm A}^{\text{(ev)}}(\bm{k}) \ne \pm {\bm A}^{\text{(ev)}}(\bm{k}),
\eea
but it is a twisted state in a sense of Eq. (\ref{jzAk}) for any $\theta$. Equation (\ref{Apar}) demonstrates that the radial quantum number $s$ stays the same, $s' = s$, within the paraxial regime, i.e., for small transverse momenta, and only the electron angular momentum changes. Thus, the emission of photons with no vorticity ($j_z = 0$) due to the transitions $s\to s'\ne s, \ell'= \ell$ occurs only beyond the paraxial approximation, i.e. with large transverse momenta. To quantitatively define the paraxiality condition, we recall that the $S$-matrix transition amplitude is proportional to the exponential function $\exp\{-(\kappa/\kappa_c)^2/2\}$ [see Eq. (\ref{expk}) and recall that $k_{\perp}=\kappa$]. This implies that transverse momenta much higher than
\bea
& \displaystyle \kappa_c = 2/\rho_H = m\sqrt{\frac{H}{H_c}}
\label{kplim}
\eea
are suppressed. The numerical value of this scale is $\kappa_c \sim 10$ eV for $H \sim 1$ T. The strong inequality 
\bea
\kappa \ll \kappa_c
\label{parax}
\eea
can be called the paraxiality condition under which it is mostly $s'=s$ and twisted photons are predominantly emitted. Likewise, non-paraxial photons with the transverse momenta $\kappa \sim \kappa_c$ may not be twisted, in the sense that they can be due to the transitions $s'\ne s, \ell = \ell'$. Importantly, if the overall number of emitted photons is large, there may still be a noticeable number of untwisted photons, exactly as photons can be emitted with energies larger than the critical energy of synchrotron radiation \cite{ST}. 

\section{Emission probability and intensity}

Having derived the state in which the photon has been emitted itself, we now investigate the radiation probability. The first-order emission probability that the photon is detected with the momentum between $\bm{k}$ and $\bm{k}+d\bm{k}$ and the electron with quantum numbers $\ell'$ and $s'$ and with longitudinal momentum between $p'_z$ and $p'_z+dp'_z$ is given by
\bea
dW^{(1)}_{s',\ell'}(p'_z,\bm{k}) = \sum\limits_{\lambda = \pm 1}|S_{fi}^{(1)}|^2 d\nu,
\eea
where
\bea
d\nu = \frac{L}{2\pi} dp'_z\, \frac{1}{2\omega}\frac{1}{(2\pi)^3} d^3 \bm{k}.
\label{Nu}
\eea
Likewise, we define the first-order differential radiation intensity by multiplying the probability by the photon energy $\omega$:
\bea
dI^{(1)}_{s',\ell'}(p'_z,\bm{k}) = \omega \sum\limits_{\lambda = \pm 1}|S_{fi}^{(1)}|^2 d\nu.
\label{Int}
\eea
The total probability is given by
\bea
W^{(1)} = \sum\limits_{s',\ell'} \int dW^{(1)}_{s',\ell'}(p'_z,\bm{k}).
\eea

When squaring the matrix element, we use the rule
\bea
\left(\delta (\omega + \varepsilon' - \varepsilon)\right)^2 \left(\delta (p_z - p_z' - k_z)\right)^2 \to \frac{T}{2\pi} \delta (\omega + \varepsilon' - \varepsilon) \frac{L}{2\pi} \delta (p_z - p_z' - k_z),
\label{Ssqr}
\eea
and integrate over $d^3\bm{k}$ in cylindrical coordinates. The corresponding differential emission probability per unit time is found as
\bea
& \displaystyle
d\dot{W}^{(1)}_{s',\ell'}(p'_z) = \frac{e^2}{\varepsilon\varepsilon'} \frac{2^{\ell 
+ \ell'}}{\pi \rho_H^2}\,\frac{s!s'!}{(s + \ell)!(s' + \ell')!}\, \sum\limits_{\lambda = \pm 1}\left|\sum\limits_{\sigma=0,\pm 1} i^{\sigma}\,d_{\sigma\lambda}^{(1)}\, \mathcal I_{\sigma}\right|^2 dp_z',
\label{ProbCyl}
\eea
where $\varepsilon'$, $\sin \theta = \kappa/\omega$, and $\kappa$ depend on $p_z'$. The integration over $p_z'$ is carried out over the region of the allowed values from Eq. (\ref{pzpint}). Correspondingly, it is $d\dot{I}^{(1)}_{s',\ell'}(p'_z)=\omega d\dot{W}^{(1)}_{s',\ell'}(p'_z)$.

The condition $p_{\perp}^2 > (p'_{\perp})^2$, i.e., $s + \ell > s'+ \ell'$ implies that, by setting $\ell' = \ell - \Delta \ell$, the final radial quantum number fulfills the inequality $s' \leq s + \Delta\ell - 1$. As can be seen, the processes with increase of the electron OAM, $\ell' > \ell$ (the photon TAM $j_z^{(\gamma)} = \ell - \ell' <0$), are generally allowed but they correspond to $s'\ne s$ (for $\Delta \ell = -1$, we have $s' \leq s-2$), and these transitions have much lower probability than processes with $s=s'$.

\begin{figure}
\center{\includegraphics[width=0.99\linewidth]{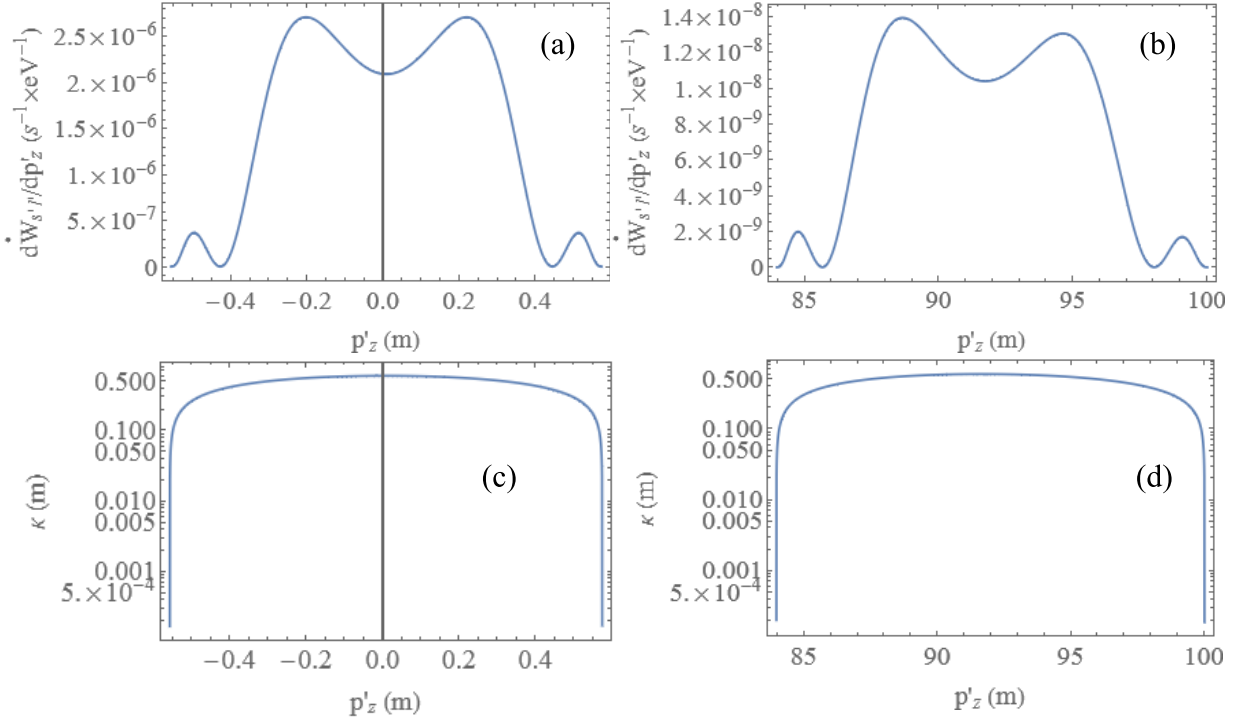}}
\caption{The differential emission probability $d\dot{W}^{(1)}_{s',\ell'}/dp'_z$ per unit time ((a), (b)) and the twisted photon transverse momentum $\kappa$ ((c), (d)) versus the final electron momentum along the field $p'_z$ for $H=H_c, s=s'=20, \ell=4,\ell'=0$. Left ((a), (c)): $p_z = 10^{-2}\,m,\, p'_{z,\text{max}}-p'_{z,\text{min}} \approx 1.13\,m$; right ((b), (d)): $p_z = 100\,m,\, p'_{z,\text{max}}-p'_{z,\text{min}} \approx 16\,m$.}
\label{Figz}
\end{figure}

Let us now estimate the photon transverse momentum $\kappa$. It generally varies from $\kappa = 0$ at the end-points (\ref{pzpint}) to $\sim\kappa_c$ from Eq. (\ref{kplim}), which reaches $\kappa_c \sim m$ at $H\sim H_c$. 
 %
 %
Figure \ref{Figz} shows that, as it can also be evinced from the expression of $p'_{z,\text{max}}$, in the presence of a magnetic field strength $H\sim H_c$, an electron with $p_z \ll m$ can in principle undergo a strong longitudinal recoil such that $|p'_z| \sim m$ after the photon emission. 

\begin{figure}
\center{\includegraphics[width=1.0\linewidth]{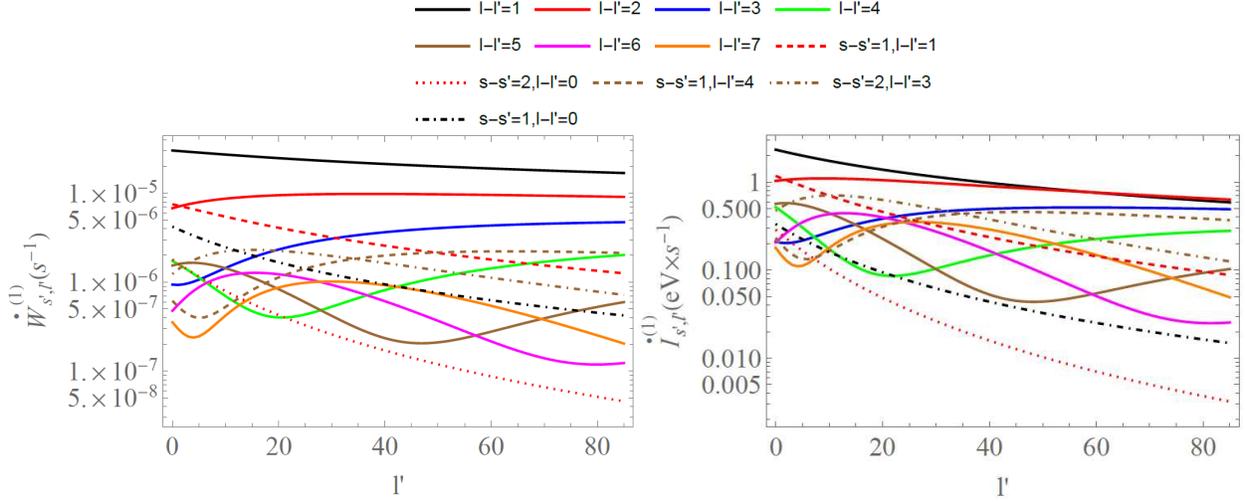}}
\caption{The emission probability (\ref{ProbCyl}) (\textit{left}) and the intensity $d\dot{I}^{(1)}_{s',\ell'}(p'_z)$ (\textit{right}) integrated over $p'_z$ per second for $H=H_c, p_z = 10^{-3}mc$. For the solid lines of the twisted photons $s=s'=20$, the dashed and dash-dotted lines correspond to the twisted photons with a simultaneous change of the radial quantum number $s\to s'\ne s$, the black dash-dotted line and the red dotted one correspond to the untwisted photons with $j_z=0$.}
\label{Fig3}
\end{figure}

As it can also be seen in Fig. \ref{Figz}, a non-relativistic electron with $p_z \ll m$ emits photons with both $k_z = p_z - p'_z >0$ and $k_z <0$, whereas with the growth of the momentum $p_z$ the electron tends to emit photons to the forward direction with $k_z = p_z - p'_{z,\text{min}} >0$. The figure also indicates that the probability is maximized far from the end-points (\ref{pzpint}) and where the photon transverse momentum reaches the value $\kappa \lesssim 0.5\, m$. One can estimate the typical transverse momenta near the end-points in the non-relativistic regime with $p_z,|p'_z|,p_{\perp},p'_{\perp}\ll m$. The first non-vanishing term in the expansion of $\kappa$ in the vicinity of $p'_z = p'_{z,\text{min}}\,\,\text{or}\,\,p'_{z,\text{max}}$ reads
\bea
& \displaystyle \kappa_{\text{non-rel}} \approx \frac{\Delta p_{\perp}^2}{2m} \sqrt{\frac{p_z^2}{m^2} + \frac{p_{\perp}^2 + (p'_{\perp})^2}{2m^2}},
\eea
with $\Delta p_{\perp}^2 = p_{\perp}^2 - (p'_{\perp})^2$. In the same approximation
\be
|k_z| \approx \frac{\Delta p_{\perp}^2}{2m} = m \frac{H}{H_c} (s + \ell -s'- \ell')  \ll m,
\label{kperp1}
\ee
and recall that the emission probability features the exponential decay $\exp\{-(\kappa/\kappa_c)^2/2\}$ defining the region $\kappa_{\text{non-rel}} \ll |k_z| \sim \kappa_c = m\sqrt{H/H_c}$, where the probability is not exponentially suppressed. In the case $s'=s$, one can express this inequality in terms of the quantum numbers $s$ and $\ell$ as follows:
\bea
& \displaystyle j_z^{(\gamma)} = \ell - \ell' \sim \sqrt{\frac{H_c}{H}},
\label{upb1}
\eea
which means that mostly the twisted photons with the moderate values of the TAM $\ell -\ell' \gtrsim 1$ are emitted at $H\sim H_c$, see Fig. \ref{Fig3}. 

\begin{figure}
\center{\includegraphics[width=1.0\linewidth]{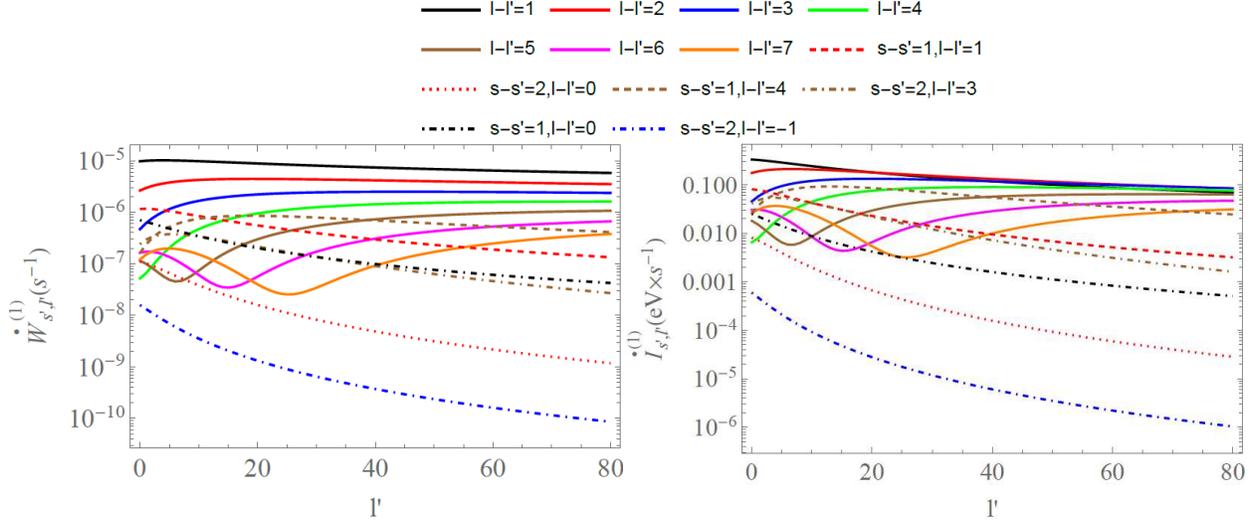}}
\caption{The emission probability (\ref{ProbCyl}) (\textit{left}) and the intensity $d\dot{I}^{(1)}_{s',\ell'}(p'_z)$ (\textit{right}) integrated over $p'_z$ per second for $H=0.1\, H_c, p_z = 10^{-3}mc$. For the solid lines of the twisted photons $s=s'=5$, the dashed and dash-dotted lines correspond to the twisted photons with a simultaneous change of the radial quantum number $s\to s'\ne s$, the black dash-dotted line and the red dotted one correspond to the untwisted photons with $j_z=0$, and the blue dash-dotted line corresponds to increase of the electron OAM during the emission (so that the photon TAM is $\ell-\ell' = -1$).}
\label{Fig4}
\end{figure}

\begin{figure}
\center{\includegraphics[width=1.0\linewidth]{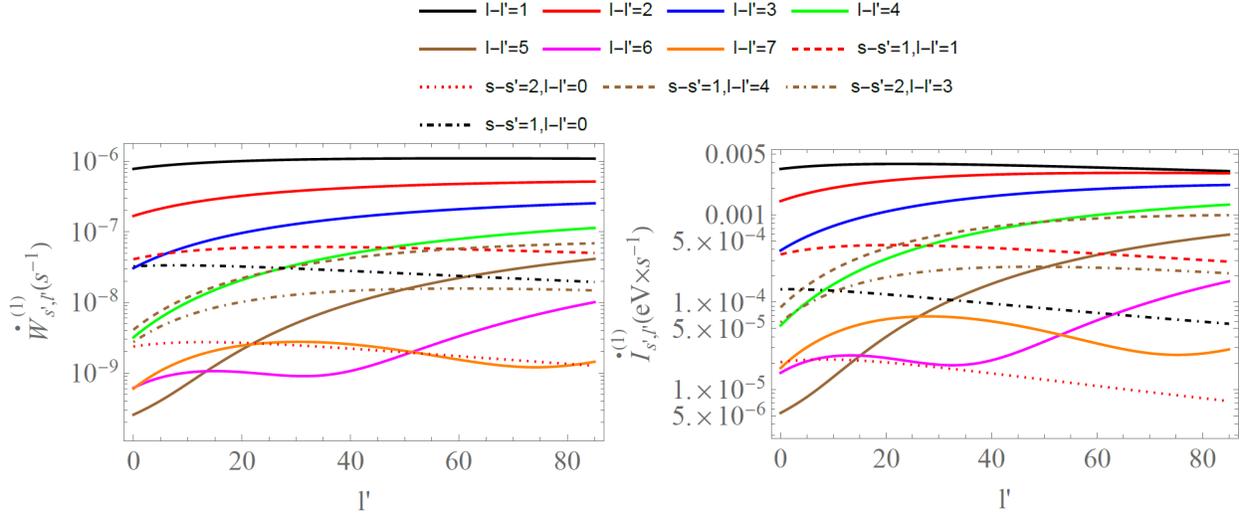}}
\caption{The same as in Fig.\ref{Fig3}, but for $H = 10^{-2} H_c$.}
\label{Fig5}
\end{figure}

Due to dependence of the radiation frequency on $\ell$ and $\ell'$ (see Figs. \ref{FigSpectrumA} and \ref{FigSpectrumB}), the emission probability and the intensity depend slightly differently on $\ell'$ for the transitions $\ell\to\ell'\ne\ell$, as shown in Fig.\,\ref{Fig3}, Fig.\,\ref{Fig4}, and Fig.\,\ref{Fig5} for a non-relativistic electron with $p_z = 10^{-3}\,m$ and for $H = H_c, H = 10^{-1}\,H_c$, and $H = 10^{-2}\,H_c$, respectively. Whereas the probability of the transitions $\ell \to \ell'=\ell-1$ always dominates, the intensity for the transition $\ell \to \ell'=\ell-2$ becomes slightly larger than that for $\ell \to \ell'=\ell-1$ starting from $\ell' \gtrsim 60$ at $H \sim H_c$. As can be seen in Fig. \ref{Fig5}, the weaker fields $H \ll H_c$ favor transition from the higher Landau levels, $\ell,\ell'\gg 1$, however, with the same typical TAM $\ell - \ell' \gtrsim 1$ and much lower intensity. Interestingly, both the probability and the intensity for the untwisted photons with $\ell'=\ell$ can be higher than those for the highly twisted photons with $\ell-\ell'\gg 1$, especially for the moderate value of $\ell'$.

\begin{figure}
\center{\includegraphics[width=1.0\linewidth]{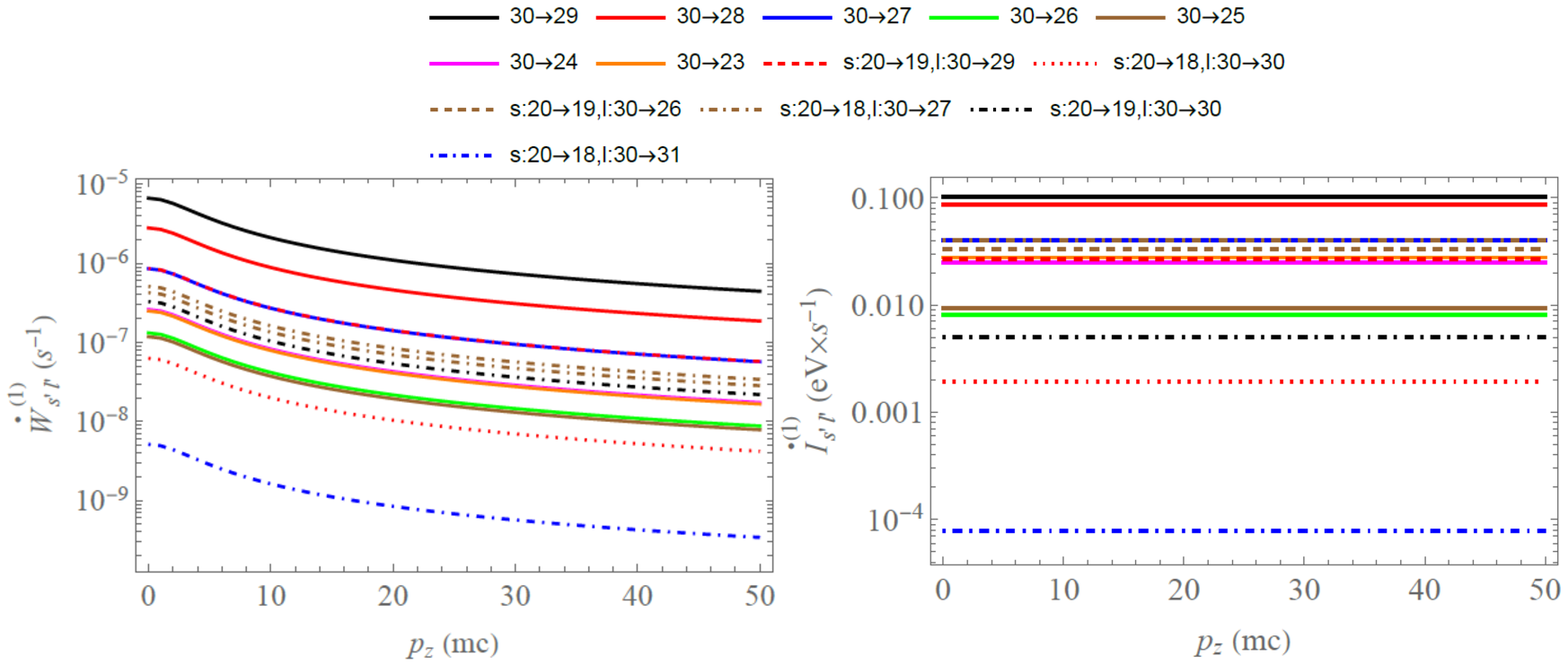}}
\caption{The dependence of the emission probability (left) and the intensity (right) per second on the electron momentum $p_z$ for $H=0.1\, H_c, s=s'=20$. The transition $30\to 29$ means $\ell = 30, \ell'=29, s=s'=20$; those with $\ell: 30\to 30$ correspond to the untwisted photons with $j_z^{(\gamma)}=0$. 
The blue line overlaps with the red dashed one on the left and with the brown dash-dotted one on the right. The blue dash-dotted lines correspond to increase of the electron OAM during the emission (so that the photon TAM is $\ell-\ell' = -1$).}
\label{Fig6}
\end{figure}

As can be seen in Fig. \ref{Figz} (b) and (d), relativistic considerations imply that the photon is much probably emitted with $k_z>0$ and such that $p'_z>0$, which we assume below, but the photon transverse momentum practically does not grow with $p_z$. 
%
%
Fig. \ref{Fig6} shows the total radiation probability and the intensity per second for the definite transverse momentum of the final electron as a function of the initial momentum $p_z$ (assumed to be positive, $p_z>0$) along the magnetic field. Although the momentum-dependent probabilities for the untwisted photons are not necessarily the lowest ones, the intensities do not depend on the momentum $p_z$ at all and demonstrate the domination of the twisted radiation with the TAM $\ell-\ell' \sim 1, 2$ and unchanged $s$. This constant behavior is related to the fact that the intensity of radiation is a Lorentz-invariant quantity (recall, for example, the expression of the classical relativistic Larmor formula). Here, the intensity of radiation is computed at fixed angular-momentum quantum number and its invariance is restricted to boosts along the magnetic-field direction, i.e., the $z$-direction. A similar situation occurs in the emission of radiation in a constant electric field \cite{Nikishov_1969}, where the independence of the spectrum from the longitudinal momentum of the electron can also be interpreted as the independence from the time origin. 

These figures illustrate that emission of the twisted photons with $\ell-\ell' \gtrsim 1$ can be quite intense in the critical and subcritical fields, and the untwisted photons due to the transitions $s\to s'\ne s, \ell=\ell'$ can also be emitted, though with lower intensity. As argued above, the twisted photons due to the general transitions $s\to s'\ne s, \ell\to\ell'\ne \ell$ have much larger transverse momenta but also much lower intensity. Moreover, we see in Fig. \ref{Figz} that electrons in the field strength $H \sim H_c$ can also emit \textit{non-paraxial} twisted photons with $\kappa \lesssim m$ whatever their longitudinal momentum $p_z$ is. The radius of their Bessel ring is of the order of \cite{Torres, And, SerboUFN, Ivanov} 
\bea
\rho_c \sim (\ell-\ell')/\kappa \gtrsim (\ell-\ell')/m \equiv (\ell-\ell')\lambda_c, 
\eea
which can be shown by evaluating the integral in Eq. (\ref{ArEv}) (recall that $\lambda_c \approx 3.8 \times 10^{-11}\,\text{cm}$ is the electron Compton wavelength). Note that the rms-radius of the Landau state at $H\sim H_c, |\ell|\gg 1$ is (see Eq. (\ref{re2})) $\sqrt{\la\rho^2\ra} \sim \lambda_c \sqrt{|\ell|} \gg \lambda_c$, but the Bessel ring radius of such non-paraxial photons grows as $\ell-\ell'$.

Since field strengths up to values of the order of $H_c$ have been considered, the question arises whether vacuum-polarization effects due to the strong magnetic field alter the dispersion relation of the emitted photon \cite{BLP}. Including vacuum-polarization effects, however, would correspond in the language of Feynman diagrams to considering the radiative corrections arising from the insertion of the polarization operator into the emitted photon line. Since we have carried out a leading-order analysis here, all radiative corrections have been consistently neglected. Indeed, even in the case of magnetic fields of the order of $H_c$ the photon refractive index would differ from unity by terms of the order of $\alpha/45$, where $\alpha\approx 1/137$ is the fine-structure constant \cite{BLP}.

\section{Conclusion}

Although quantum dynamics of Landau electrons in a magnetic field has been studied from 1930s (see the literature, for instance, in Refs. \cite{ST, Bagrov}), it has only been recently recognized that the electron can emit twisted photons while going from one Landau level to another \cite{Mar, Kruining}. Here, we have answered the question of whether the photon emitted by a spinless relativistic charged particle is twisted, irrespective of the photon detector properties. We have found that the overwhelming majority of the emitted photons are indeed twisted, representing the Bessel beams with the TAM projection $j_z^{(\gamma)} = \ell -\ell'$, although a small part of them are emitted due to the transitions $s \to s'\ne s, \ell'=\ell$ without a change of the angular momentum. The latter photons are \textit{not} twisted, which represents yet another difference as compared to the predictions of classical electrodynamics from Refs. \cite{Katoh, Katoh2, Epp, Epp2}. 

The angular momentum of the photons turns out to be quantized along the magnetic field direction, which is why the radiation vorticity does not usually reveal itself at the storage ring facilities where the photons are being detected at the angles close to the orbital plane, $\theta \approx \pi/2$, and the wave front looks locally flat. Indeed, the radiation vorticity can be noticed when observing the emission at angles close to the magnetic field axis, $\theta \to 0$, especially for the critical and sub-critical fields, $H \lesssim H_c$, typical for neutrons stars. As it has already been emphasized \cite{Mar}, the vorticity of the resulting photons can be important for studying the stellar nucleosynthesis because twisted photons interact differently with other charged particles in the sense that -- unlike the plane-wave photons -- they can excite other Landau electrons to the states with higher angular momenta or induce stimulated emission. These differences from interactions of the plane-wave photons can be especially important for twisted photons with the moderate angular momenta $\ell - \ell' \gtrsim 1$ and the large transverse momenta $\kappa \lesssim m$ (so that $\rho_c \gtrsim \lambda_c$), generated in the high intensity fields \footnote{While finalizing the manuscript, we have become aware of the recent paper \cite{EppH} treating the very similar problem, however, without analyzing the photon evolved state.}.

\

D.K. is grateful to V.~Bagrov, D.~Grosman, V.~Serbo, A.~Surzhykov, A.~Volotka, and V.~Zaytsev for fruitful discussions and criticism, and, especially, to S.~Baturin, G.~Sizykh, and A.~Pupasov-Maksimov for interesting discussions and help with the integrals. D.K. is also grateful to the A. von Humboldt Foundation for financial support, to the Max Planck Institute for Nuclear Physics for the hospitality, and to I.\,Pavlov who pointed out a typo in the numerical code.

\appendix

\section{Radial integrals}

Equation (20.17) in \cite{ST} can be cast as follows: 
\bea
& \displaystyle \int\limits_0^{\infty} dx\, x^{\ell + \ell' + 1} L_s^{\ell}(x^2) L_{s'}^{\ell'}(x^2) J_{\ell-\ell'}(2x\sqrt{y}) e^{-x^2} = \cr 
& \displaystyle = \frac{1}{2}\,\frac{(s'+ \ell')!}{s!}\,y^{s-s'+\frac{\ell - \ell'}{2}} L_{s'+ \ell'}^{s + \ell - s'- \ell'}(y) L_{s'}^{s-s'}(y)\, e^{-y}.
\label{STs0}
\eea
Let us denote its modified version as
\bea
& \displaystyle F_{s,s'}^{\ell,\ell'}(y) = \int\limits_0^{\infty} dx\,x^{\ell+\ell'+1}\, L_{s}^{\ell}(2x^2)L_{s'}^{\ell'}(2x^2)\,J_{\ell-\ell'}(y x)\, e^{-2x^2} = \cr 
& \displaystyle = \frac{(s'+\ell')!}{s!}\frac{1}{2^{3(s-s') + 2\ell - \ell' + 2}}\, y^{2(s-s') + \ell-\ell'}\, L_{s'+\ell'}^{s-s'+\ell-\ell'}\left(y^2/8\right)L_{s'}^{s-s'}\left(y^2/8\right)\, e^{-y^2/8}.
& \displaystyle 
\label{F}
\eea

We start from the expression for the radial integral in Eq. (\ref{Is}) that reads 
 \bea
& \displaystyle \mathcal I_{-1} =-i\sqrt{2}\int\limits_0^{\infty} dx\, x^{\ell+\ell'+2}\, J_{\ell-\ell'+1}(y x) \, e^{-2x^2}\cr
&\times\left[2L_{s-1}^{\ell+1}L_{s'}^{\ell'} - 2L_{s}^{\ell}L_{s'-1}^{\ell'+1}+\frac{\ell'}{x^2}L_{s}^{\ell}L_{s'}^{\ell'}+2L_{s}^{\ell}L_{s'}^{\ell'}\right].
  \label{eq:Imm}
 \eea
By using the following relations for Laguerre polynomials
\begin{align}
    L_s^l(x)=L_s^{\ell+1}(x)-L_{s-1}^{\ell+1}(x),
    \label{eq:rec0}
\end{align}
and
\bea
    & \displaystyle xL_s^\ell(x)=(\ell+s)L_{s}^{\ell-1}(x)-(s+1)L_{s+1}^{\ell-1}(x),
    \label{eq:rec1}
\eea
we find
\bea
& \displaystyle 2x^2 L_{s-1}^{\ell+1}L_{s'}^{\ell'} - 2x^2 L_{s}^{\ell}L_{s'-1}^{\ell'+1} + \ell' L_{s}^{\ell}L_{s'}^{\ell'}+2x^2 L_{s}^{\ell}L_{s'}^{\ell'} = 2x^2 L_s^{\ell+1} L_{s'}^{\ell'} + (s' + \ell') L_s^{\ell} L_{s'}^{\ell'-1}.
\eea
and then
 \bea
& \displaystyle \mathcal I_{-1} =-i\sqrt{2}\left(2 \mathcal F_{s,s'}^{\ell+1,\ell'} + (s'+\ell') \mathcal F_{s,s'}^{\ell,\ell'-1}\right).
 \eea
Analogously for $\sigma = +1$ we have
 \bea
& \displaystyle \mathcal I_{+1} =i\sqrt{2}\int\limits_0^{\infty} x^{\ell+\ell'+2}\, J_{\ell-\ell'-1}(y x) \, e^{-2x^2}\left[2L_{s-1}^{\ell+1}L_{s'}^{\ell'} - 2L_{s}^{\ell}L_{s'-1}^{\ell'+1}-\frac{\ell}{x^2}L_{s}^{\ell}L_{s'}^{\ell'}-2L_{s}^{\ell}L_{s'}^{\ell'}\right]\, dx.
  \label{eq:Ipm}
 \eea
We apply the following relations
\bea
& \displaystyle 2x^2 L_{s-1}^{\ell+1}L_{s'}^{\ell'} - 2x^2 L_{s}^{\ell}L_{s'-1}^{\ell'+1} - \ell L_{s}^{\ell}L_{s'}^{\ell'}-2x^2 L_{s}^{\ell}L_{s'}^{\ell'} = -2x^2 L_s^{\ell} L_{s'}^{\ell'+1} - (s + \ell) L_s^{\ell-1} L_{s'}^{\ell'},
\eea
and finally get
 \bea
&\mathcal I_{+1} =-i\sqrt{2}\left(2 \mathcal F_{s,s'}^{\ell,\ell'+1} + (s+\ell) \mathcal F_{s,s'}^{\ell-1,\ell'}\right).
 \eea

\end{document}